\documentclass[pss]{wiley2sp} 
\usepackage{graphicx}
\usepackage{amsmath,amsfonts,amssymb}

\begin{document}

\title{Characterization and control of charge transfer in a tunnel junction}

\titlerunning{Characterization and control of charge transfer in a tunnel junction}


\author{%
  Julien Gabelli\textsuperscript{\Ast,\textsf{\bfseries 1}},
  Karl Thibault \textsuperscript{\textsf{\bfseries 2}},
  Gabriel Gasse \textsuperscript{\textsf{\bfseries 2}},
  Christian Lupien \textsuperscript{\textsf{\bfseries 2}},
  Bertrand Reulet \textsuperscript{\textsf{\bfseries 1,2}},}

\authorrunning{Gabelli et al.}

\mail{e-mail
  \textsf{julien.gabelli@u-psud.fr}, Phone:
  +33-169155365, Fax: +33-169155025}

\institute{%
  \textsuperscript{1}\,Laboratoire de Physique des Solides, CNRS, Univ. Paris-Sud, Universit\'{e} Paris-Saclay, 91405 Orsay Cedex, France\\
  \textsuperscript{2}\,D\'{e}partement de Physique, Universit\'{e} de Sherbrooke, Sherbrooke, Qu\'{e}bec J1K 2R1, Canada}

\received{XXXX, revised XXXX, accepted XXXX} 
\published{XXXX} 


\abstract{%
%
%
%
\abstcol{Charge transfer in a tunnel junction is studied under dc and ac voltage bias using quantum shot noise. Under dc voltage bias $V$, spectral density of noise measured within a very large bandwidth enables to deduce the current-current correlator in the time domain by Fourier transform. This correlator exhibits regular oscillations proving that electrons try to cross the junction regularly, every $h/eV$.  Using harmonic and bi-harmonic ac voltage bias, we then show that quasiparticles excitations can be transferred through the junction in a controlled way. By measuring the reduction of the excess shot noise, we are able to determine the number of electron-hole pairs surrounding the injected electrons and demonstrate that bi-harmonic voltage pulses realize an on-demand electron source with a very small admixture of electron-hole pairs. }}

%
%
\titlefigure[height=3cm]{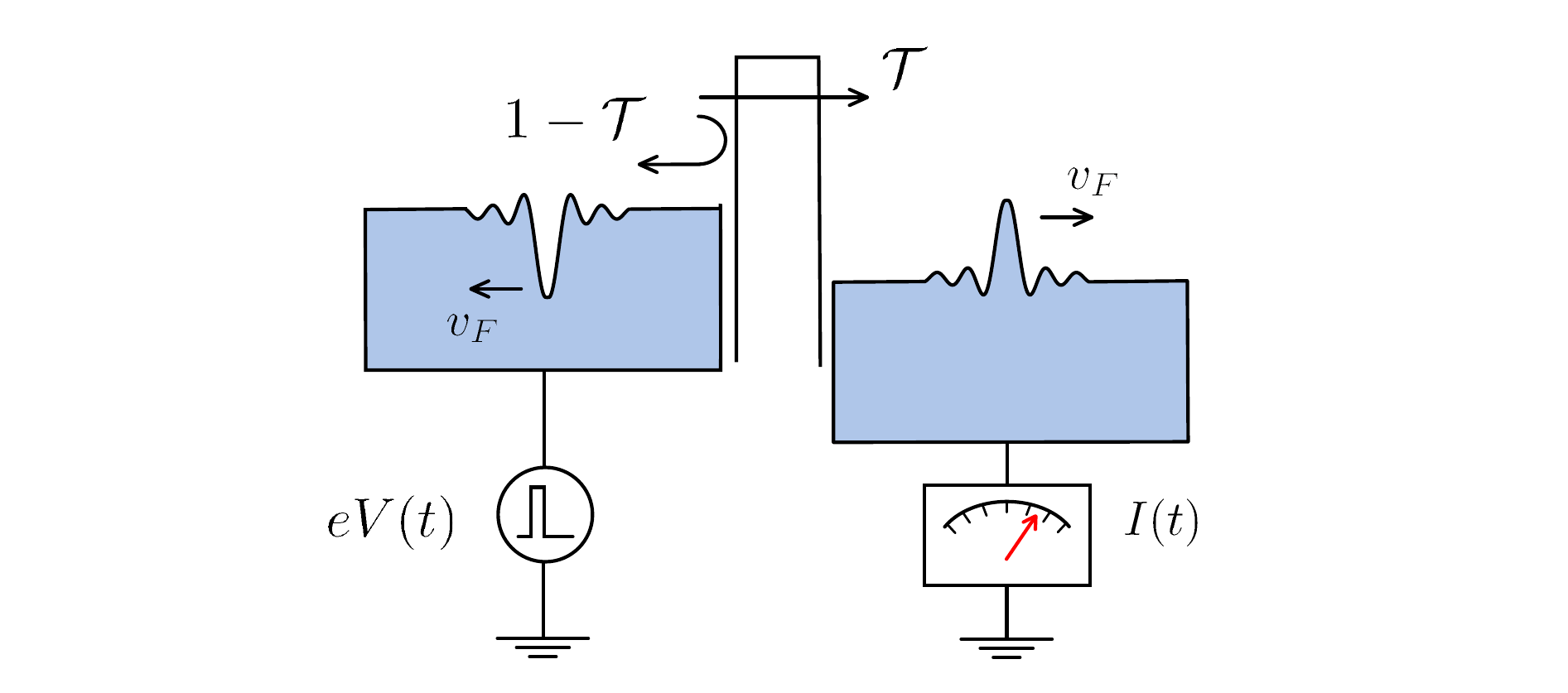}
\titlefigurecaption{%
  A time-dependent voltage drive $V(t)=V_{dc}+V_{ac}(t)$ applied to a contact with transmission $\mathcal{T}$ generates an incoming excitation giving rise to transmitted and reflected quasiparticles. The excess noise $\Delta S_{V}=(\langle I(t)^2 \rangle_{ac+dc}-\langle I(t)^2 \rangle_{dc})/\Delta f$ given by the difference between the noise measured with and without the ac excitation is measured by an ammeter with a bandwidth $\Delta f$. It  gives the number of electron-hole pairs surrounding the transmitted electrons: $N_{e-h}=\frac{h}{2e^2\mathcal{T}} \frac{\Delta S_{V}}{h\nu}$ where $\nu$ is the repetition frequency of $V_{ac}$.}

\maketitle   

\section{Introduction.}

Low temperature electron transport in nanostructures reveals the wave nature of electron propagation in conductors and is at the heart of the emerging field of electron quantum optics. Since the early 90s, high mobility two-dimensional electron  gases (2DEG) succeed in realizing electronic analogs of all kind of optical interferometers such as double slit \cite{Yacoby94,Schuster97}, Fabry-P\'{e}rot \cite{Chamon97} or Mach–Zehnder interferometers \cite{Ji03}. However, if one refers to optics, all these interferometers deal with one-photon interferences and can be described by the classical theory of light. Quantum optics effects are indeed revealed by a two-particle correlation measurement \cite{Glauber63} and this is why the Hanbury Brown and Twiss correlation measurement \cite{HBT56} realized by Kimble \textit{et al.} \cite{Kimble77} with a single photon source is usually considered as a landmark in the field of quantum optics. In this experiment, the observation of photon antibunching in light emitted by a single atom unambiguously demonstrated the non-classical nature of the single photon source. In electron quantum optics, this step was taken more than 15 years before the implementation of the single-electron source  \cite{Feve07,Dubois13,Feve13,Jullien14,Feve16} when Oliver \textit{et al.} \cite{Oliver99} and Henny \textit{et al.} \cite{Henny99} realized independently an HBT interferometer in a 2DEG. In these experiments, electrons were emitted by a dc voltage biased metallic contact and antibunched due to the fermionic statistics without engineering a single-electron source. It demonstrates the fundamental distinction between electron and photon quantum optics: the difference in vacuum state. Unlike the photon vacuum, the electronic vacuum - the Fermi sea - can be controlled by a voltage bias. This property, together with the quantum correlations induced by the Pauli principle, are at the heart of single electron sources in electron quantum optics. Our ability to realize a single electron source seems to be independent of the conductor in which electrons are injected and only depends on the metallic contact on which the voltage bias is applied.
In this article, we focus on the experimental realization of electron sources using voltage pulses to create elementary excitations from a degenerate Fermi sea in the contacts of a conductor. In this context, injected quasiparticles are electrons surrounded by an electron-hole cloud whose number and probability of creation depend on the shape and the amplitude of the applied voltage. It has been demonstrated that quantized Lorentzian pulses $V(t)$ such as $\int eV(t)/h \, dt = N$ ($N$ is integer) create a "pure" excitation of $N$ electrons above the Fermi sea at zero temperature \cite{Levitov96,Levitov97}. Repeated injection of electrons thus requires a periodic signal with infinite number of harmonics, which is experimentally not feasible. The aim of this study is to experimentally tailor the voltage pulses with a bi-harmonic signal in order to minimize the electron-hole cloud compared to the transferred electrons between two metallic contacts toward the implementation of single electron source. We show that a purity of less better than 99\% can be achieved for the injection of a single electron with bi-harmonic excitation (as compared to $\sim$97\% with one harmonic only). We also discuss the effect of finite temperature on this purity.

Even if the average number of transferred electrons is governed by the conductor between the contacts, the nature of the excitation depends only on the preparation of the coherent state in the contacts. In our case, we have chosen the simplest coherent conductor, a tunnel junction, taking advantage of the significant shot noise provided by a low transmission barrier. The presence of electron-hole pairs in the injected quasiparticles can indeed be probed by the current noise power which is increased with respect to the dc noise level. Before describing the noise spectroscopy of electron-hole pairs in Section 3, we first study the quantum correlations in a coherent metallic contact at very low temperature with the degenerate Fermi sea under dc voltage bias in Section 2.

\section{Correlations in a coherent metallic contact.}

\vspace{0.2cm}
\noindent In a typical quantum transport experiment, a nanostructure is placed at low temperature between two massive metallic contacts enabling electrical measurements with macroscopic apparatus. The quantum transport in the sample is perfectly described by a very simple idea: when the conductor is voltage biased, contacts emit electrons which are either transmitted or reflected. If quantum mechanics provides the transmission/reflection probabilities, it also induces strong correlations between successive attempts of the electrons to cross the sample. Indeed, the wave-packet picture introduced by Martin and Landauer \cite{Martin1992} tells us that Fermi statistics is responsible for the regular emission of electrons and the absence of electronic noise for a perfectly transmitting conductor. The constant voltage source acts as a single electron turnstile and the average time between electrons emitted from the macroscopic contact is $h/eV$. This time does not depend on the nature of the conductor but only on the voltage drop $V$ over the conductor and fundamental constants -- the elementary charge $e$ and the Planck constant $h$.  A key question must be elucidated concerning this regular quantum electron source: what is defining its coherence? Here we show experimentally how temperature is the only parameter which defines the electron source coherence. A current-current correlator characterizing the coherence of the voltage source is deduced from the study of the current fluctuations $i(t)$ in a tunnel junction. This correlator is obtained from the spectroscopy of the noise emitted by the junction within a very broad frequency range and exhibits a simple form: $\Delta C(t,T,V)=\Delta C_{eq}(t,T) \, \cos \left(eVt/\hbar \right)$ with $\Delta C_{eq}$ the current-current correlator at equilibrium, in the absence of voltage bias (a precise definition of the correlators is given below). While the temperature $T$ leads to a jitter which tends to decorrelate electron transport after a time $\hbar/k_BT$, the bias voltage $V$ induces strong correlations which oscillate with a period $h/eV$.

\begin{figure}[t]
\includegraphics[width=\linewidth]{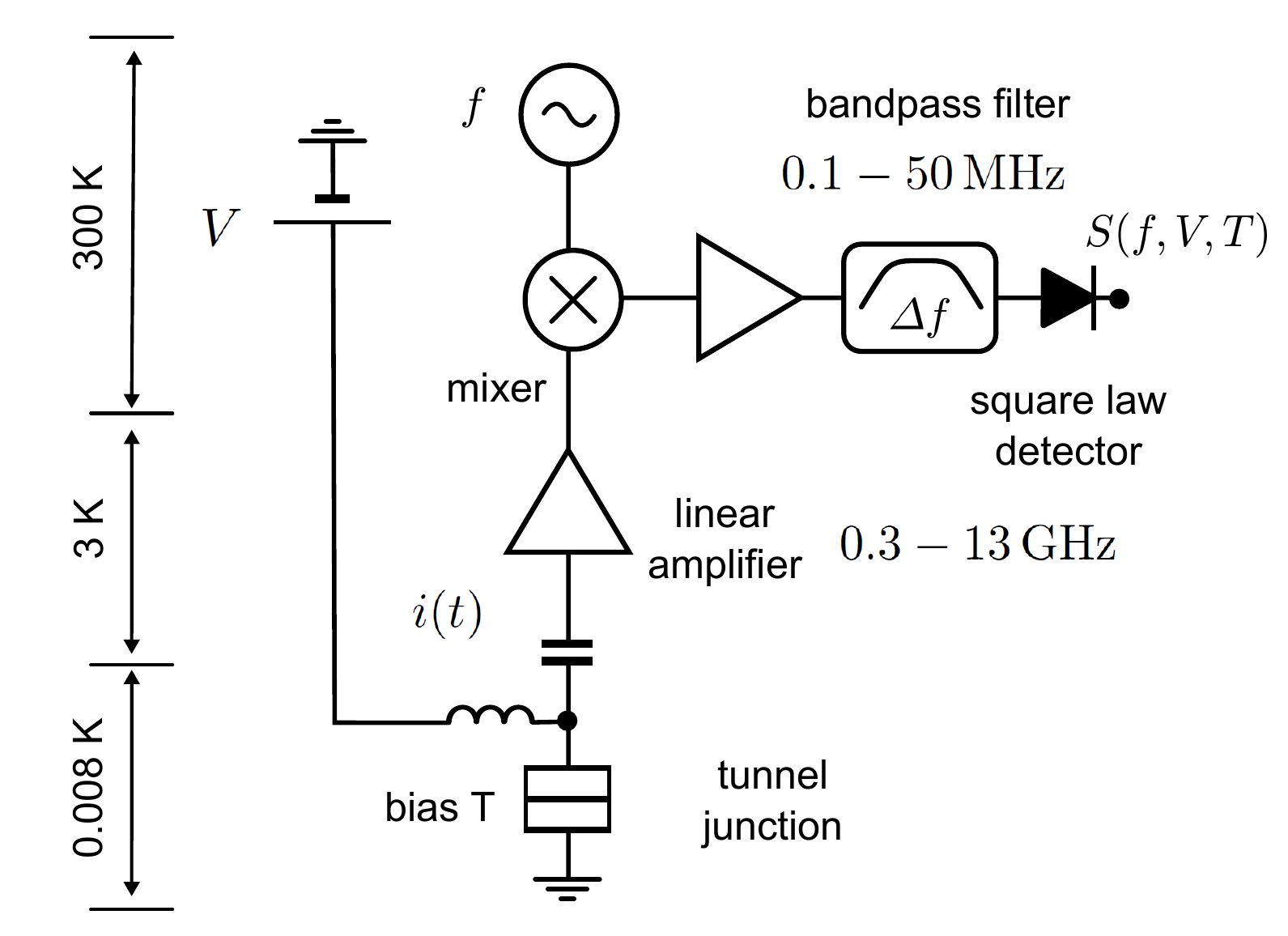}
\caption{Experimental setup used for the spectroscopy of the noise emitted by a tunnel junction.}
\label{fig_setup1}
\end{figure}

\subsection{Samples and experimental setup.}

\vspace{0.2cm}
\noindent The  Al/Al oxide/Al tunnel junctions used in the three experiments described in this article have been fabricated with the same techniques and slightly differ by their geometry (from $5$ to $10$~$\mu \mathrm{m}^2$) and their resistance (from $48$ to $70$~$\Omega$).
They have been fabricated by photolithography using the Dolan bridge technique \cite{Dolan1977}. These junctions have been then placed on the cold plate of a dilution refrigerator, which temperature can be adjusted above its base value, $\sim 10$~mK, with a resistive heater. A $\sim 500$~Gauss perpendicular magnetic field kept the Al in its non-superconducting state, with negligible effect on charge transport.

In the first experiment, that of ref.~\cite{Thibault2015}, the resistance of the junction $R=51$ $\Omega$ is voltage- and temperature-independent within less than 1\% in all measurements. The detection setup, depicted in Fig.~\ref{fig_setup1}, is similar to that of ref.~\cite{Schoelkopf1997}. The dc bias is applied to the junction through the dc port of a bias-tee. The high frequency current fluctuations generated by the sample are amplified at 3~K by a high-electron-mobility transistor 0.3-13~GHz amplifier. The resulting signal is, after further room temperature amplification, down converted by multiplication with a local oscillator of variable frequency $f$. We measure the power of the low frequency resulting signal after filtering between 0.1-50~MHz, from which we deduce the noise spectral density of the sample $S(f)$. Our calibration procedure is detailed in \cite{Thibault2015}. $S(f)$ is given by $S(f)=\langle |i(f)|^2\rangle \Delta f$ where $i(f)$ is the Fourier component of the fluctuating current $i(t)$ taken at frequency $f$, and where $\Delta f$ is the bandwidth of the measurement. The average $\langle . \rangle$, which corresponds to statistical average in theory, is performed experimentally by averaging over a time much longer than $1/\Delta f$. It is convenient to express the noise by an equivalent temperature $T_N(f) = S(f)/(2k_BG)$, where $G$ is the conductance of the sample, so we will present some experimental results in terms of $T_N$.

Fitting the measured low frequency ($hf \ll k_BT$) noise spectral density using \cite{Dahm1969}:
\begin{equation}
S(f=0,V,T)=GeV\coth (eV/2k_BT),
\label{eq:S(V,f=0)}
\end{equation}
we can extract the sample's electron temperature \cite{Spietz2003}. At the coldest point of the refrigerator, where the phonon temperature is $T_{ph}=8$~mK, we obtain an electron temperature of $T=35$~mK. This difference is most likely caused by the cold amplifier emitting noise with very wide bandwidth towards the sample, thus heating the electrons.
\begin{figure}[t]
\begin{center}
\includegraphics*[width=0.9\linewidth]{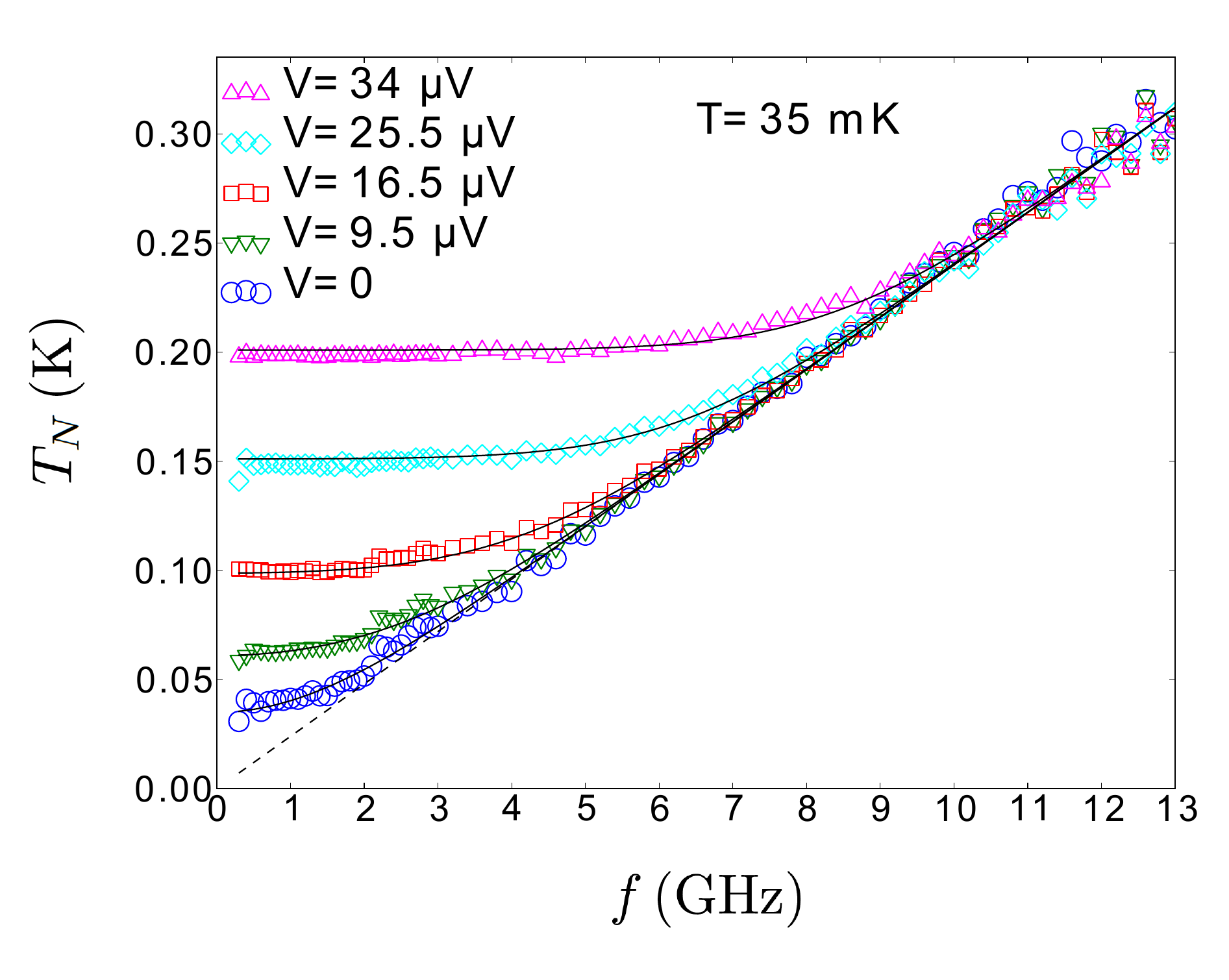}
\end{center}
\caption[]{Out of equilibrium noise temperature vs. frequency for different dc voltage biases $V$ at $T = 35$~mK. Symbols are experimental data and solid lines are theoretical expectations of Eq.~(\ref{eq:S_V}). Taken from ref.~\cite{Thibault2015}.}
\label{fig:S_V}
\end{figure}

\subsection{Theoretical expectations.}
At equilibrium, the spectral density of noise for a tunnel junction is predicted to be \cite{Callen1951}:
\begin{equation}
S_{eq}(f, T) = Ghf \coth \left( \frac{hf}{2 k_B T} \right).
\label{eq:S_eq}
\end{equation}
When a dc bias is applied to the sample, this becomes \cite{Dahm1969}:
\begin{equation}
S(f, V, T) = \frac{1}{2} \left[ S_{eq}\left( f_+, T \right) + S_{eq}\left( f_-, T \right) \right],
\label{eq:S_V}
\end{equation}
\noindent where $f_{\pm}= f \pm eV/h$. This noise spectral density has a very simple form after Fourier Transform (FT): $C(t) = \int_{0}^{\infty}\frac{1}{2\pi}\cos(2\pi f t)S(f) df$. Indeed, Eq.~(\ref{eq:S_V}) leads to the current-current correlator in time domain: $C(t,T,V)  = C_{eq}(t,T) \cos (eVt/\hbar)$. However, $C_{eq}(t,T)$ is not well defined since $S_{eq}(f)$ diverges as $\vert f \vert \rightarrow \infty$. Therefore, we introduce the \textit{thermal excess noise}:
\begin{equation}
 \Delta S_{T}(f,T,V) = S(f,T,V) - S(f,T=0,V),
\label{eq:deltaST}
\end{equation}
\noindent which vanishes as $\vert f \vert \rightarrow \infty$. Its FT obeys:
\begin{equation}
\label{eq:delta_C_V}
\Delta C(t,T,V) =\Delta C_{eq}(t,T) \cos\left( \frac{eVt}{\hbar} \right),
\end{equation}
\noindent where $\Delta C_{eq}(t,T) = C_{eq}(t,T) - C_{eq}(t,0)$. Here $C_{eq}(t,0)=\textrm{FT}[S_{vac}(f)]$ corresponds to the (infinite) jitter associated with zero point fluctuations. To obtain such a simple and remarkable result, it is essential to subtract from $S(f,T,V)$ the zero temperature \emph{but finite voltage} noise spectral density, not the zero temperature, zero voltage vacuum fluctuations $S_{vac}(f) = Ghf$.

\subsection{Experimental results.}
We have measured $T_N$ vs frequency for various bias voltages $V$. The data at the lowest electron temperature are shown on Fig.~\ref{fig:S_V}. There are two interesting limits to consider. At low frequencies, $hf < eV$, there is a plateau corresponding to $S=eI$ the classical shot noise \cite{Schottky1918}. Conversely at high frequencies, $hf \gg eV$, the data collapses on the dotted line $S=S_{vac}(f)$ given by the vacuum fluctuations. This collapse is a result of our measurement and not an hypothesis. Our only assumption is that the noise at high voltage is given by $S(eV \gg hf , k_BT) = eI$. Black lines on Fig.~\ref{fig:S_V} are the theoretical predictions of the out of equilibrium noise spectral density given by Eq.~(\ref{eq:S_V}), with no adjustable parameters. The data are in very good agreement with the theory.

\begin{figure}[t]
\begin{center}
\includegraphics*[width=0.9\linewidth]{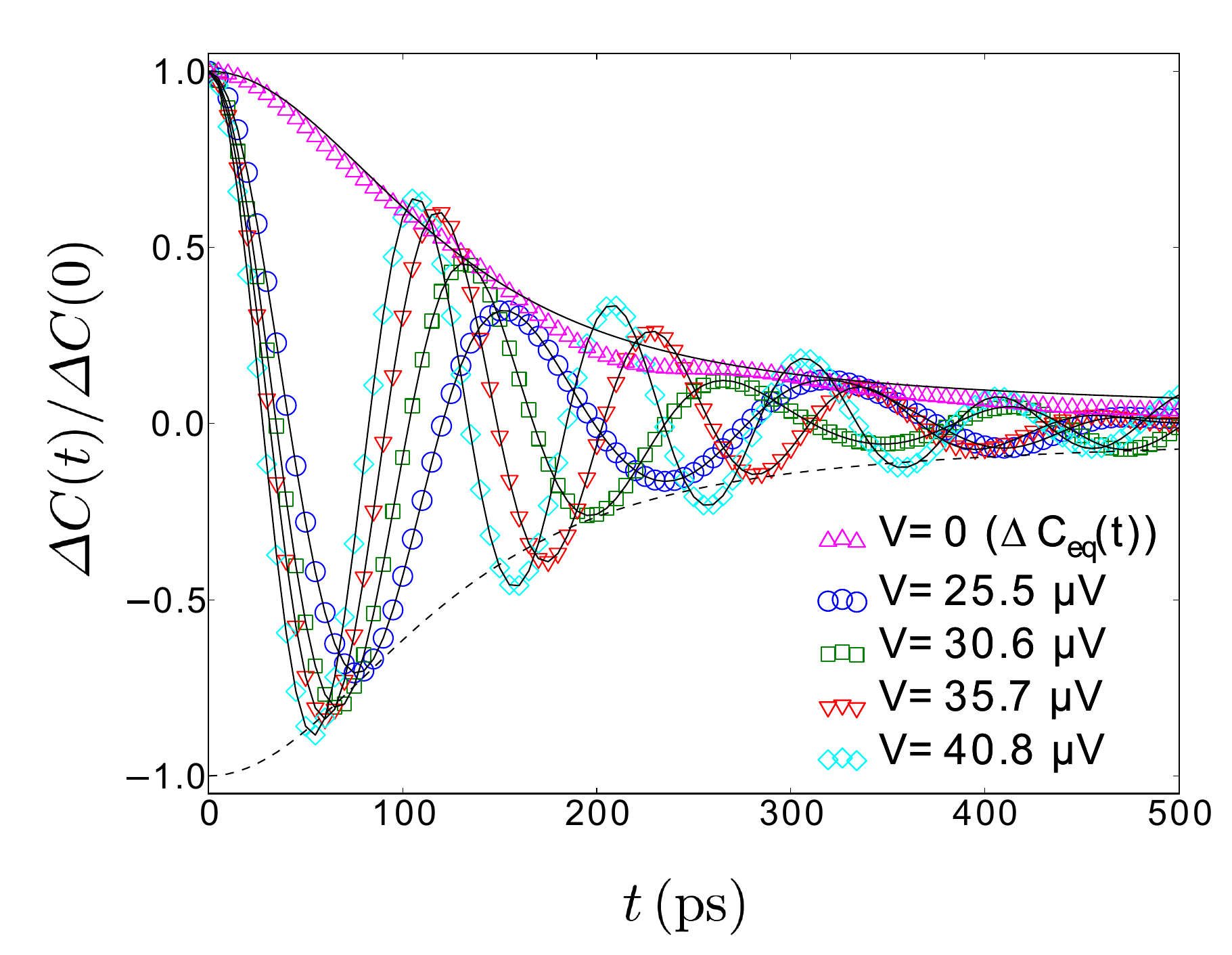}
\end{center}
\caption[]{Rescaled current-current correlator in time domain for five different voltages at $T= 35$~mK. The data at $V=0$ correspond to the correlator at equilibrium $\Delta C_{eq}(t,T)$. Its characteristic thermal decay time is given by $\hbar/k_BT \sim 100$~ps. Solid lines are theoretical expectations. Taken from ref.~\cite{Thibault2015}.}
\label{fig:pico_C(t)}
\end{figure}

Since the full spectroscopy of the spectral density of noise was measured, it is possible to use the data on Fig.~\ref{fig:S_V} to calculate the current-current correlator $\Delta C(t,T,V)$ by FT. The result is plotted on Fig.~\ref{fig:pico_C(t)}. We observe the equilibrium correlator $\Delta C_{eq}(t) = \Delta C(t, T= 35~\mathrm{mK}, V=0)$ (magenta symbols) to decay with a time constant given by $h/k_BT$ of $\sim 100$ ps for $T = 35$~mK. Moreover, Fig.~\ref{fig:pico_C(t)} also illustrates that the non-equilibrium correlator $\Delta C(t,T,V)$ at $T = 35$~mK clearly oscillates within the enveloppe given by $\Delta C_{eq}(t)$ with a period that depends on the bias voltage as $h/eV$, in agreement with Eq.~(\ref{eq:delta_C_V}). The voltage dependence of the oscillation period is distinctly established when the data is plotted as a function of the rescaled time $h/eV$ (see Fig.~5 of ref.~\cite{Thibault2015}).

\subsection{Discussion.}
\vspace{\baselineskip}

The oscillations in $\Delta C(t)$ are the result of the interplay between the Pauli principle and Heisenberg uncertainty principle. Indeed, suppose two electrons of energy $E$ and $E'$ cross a single channel conductor at time $t=0$. According to the Pauli principle, their energies must be different. How different they should be is given by the Heisenberg uncertainty principle: it takes a time $t_H \simeq h/(\vert E-E' \vert )$ to resolve the two energies, so $E$ and $E'$ cannot be considered different for times shorter than $t_H$. Thus, both electrons cannot cross the sample at $t=0$, the second one must wait a time $t_H$. When the sample is dc biased, $|E-E'|<eV$ thus $t_H > h/eV$. So there is a minimum time lag $h/eV$ between successive electrons. Hence, at high bias voltage, $eV \gg k_BT, hf$, the time lag $t_H$ becomes so small that electrons are basically independent. In this limit, we recover the Poisson statistics of tunneling electrons with $S=eI$. Conversely, at low bias voltage, successive tunneling electrons are correlated due to the time lag being finite and the resulting current distribution is no longer Poissonian. The regular oscillations of $\Delta C(t)$ are a direct consequence of the induced correlations between electrons due to the fact that they try to cross the sample at a pace of one electron per channel per spin direction every $h/eV$. The decay at long time that we observe on $\Delta C(t)$ is a consequence of a jitter of pure thermal origin.

Our sample, the tunnel junction, is a special case in which all conduction channels have low transmission and the Fano factor $F =1$. In the general case,  Eq.~(\ref{eq:delta_C_V}) is replaced by:
\begin{equation}
\Delta C(t) = F \Delta C_{eq}(t)\cos \left( \frac{eVt}{\hbar} \right)+ (1-F) \Delta C_{eq}(t). \nonumber
\end{equation}
For a perfect conductor, $F=0$ and $\Delta C(t) = \Delta C_{eq}(t)$. Hence, a perfect conductor would exhibit no oscillation of the current-current correlator because there is no shot noise \cite{Reznikov1995,Kumar1996}.

\section{Dynamical control of the charge transfer with a pulsed voltage source.}

\vspace{0.2cm}
\noindent After studying the quantum correlations in a coherent metallic contact at very low temperature under dc voltage bias, we focus on the dynamical voltage control of the  charge transfer with a pulsed drive $V(t)=V_{dc}+V_{ac}(t)$. Since the voltage appears to modulate the frequency at which electrons try to cross the sample, one should be able to control how electrons are transferred through the sample by controlling the time dependence of $V(t)$. A measure of this transfer, in particular the amount of electron-hole pairs created per injected electron, is given at zero temperature by zero frequency noise measurements \cite{Levitov96,Klich,Degiovanni,Ambrumenil,Vanevic}. Here the relevant quantity is the \emph{ac voltage excess noise}:
\begin{equation}
\Delta S_{V}=S(f=0,T,V_{dc},V_{ac})-S(f=0,T,V_{dc},V_{ac}=0),
\label{eq:deltaSV}
\end{equation}
\noindent with $S(f=0,T,V_{dc},V_{ac})$ the photon-assisted noise spectral density measured at low frequency ($hf \ll k_BT$). It  is related to the equilibrium noise spectral density $S_{eq}(f,T)$ by \cite{Lesovik1994}:
\begin{equation}
S(0,T,V_{dc},V_{ac})=\sum_{n=-\infty}^{+\infty} \left|c_n \right|^2 \, S_{eq}(eV_{dc}/h+n\nu,T),
\label{eq5}
\end{equation}
\noindent with $c_n$  the Fourier coefficients of $\exp \left( -\frac{i}{\hbar} \int_0^t eV_{ac}(t') \, dt' \right)$. The periodic ac voltage $V_{ac}(t)$ at frequency $\nu$ induces photon-assisted processes in the electron transport. The voltage excess noise $\Delta S_V$ measures the contribution of these processes to the noise generated by the sample.
At zero temperature, the number $N_{e-h}$ of electron-hole excitations surrounding the $N_e$ transferred charges during a cycle is related to $\Delta S_V$ by:
\begin{equation}
\label{eq:Neh}
N_{e-h}=\frac{\Delta S_V}{G h\nu}.
\end{equation}

\begin{figure}[t]%
\begin{center}
\includegraphics*[width=0.7\linewidth]{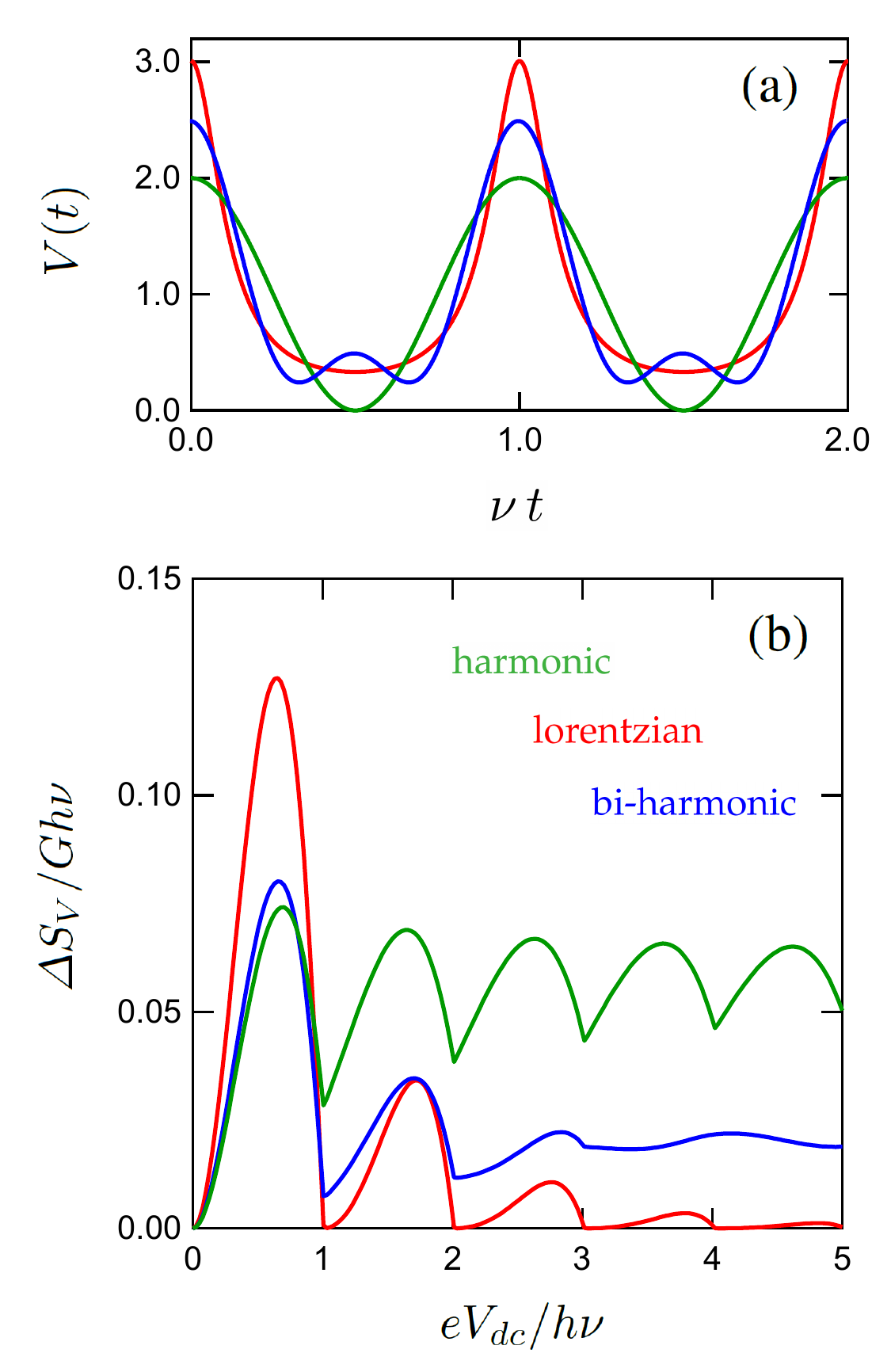}
\end{center}
\caption{(a) Periodic sequence of Lorentzian pulses of width $\tau=\mathrm{ln}\,2/2 \pi \, \nu$ (red line, $V_L(t)$) and its harmonic (green line, $V_1(t)$) and bi-harmonic (blue line, $V_2(t)$) approximations. (b) Excess noise for the different waveforms at zero temperature.}
\label{fig_DeltaSii_th}
\end{figure}

For a conductor with energy-independent transmission, the ac voltage excess noise $\Delta S_V$ goes to zero at zero temperature when the excitation is a  sequence of Lorentzian peaks of quantized area $\int_0^{\nu^{-1}} eV(t) \, dt=Nh$ each, with $N$ integer. We show on Fig.~\ref{fig_DeltaSii_th}(a) different waveforms (harmonic, biharmonic and Loretzian) and on Fig.~\ref{fig_DeltaSii_th}(b) the corresponding rescaled theoretical, zero temperature ac voltage excess noise. In the tunneling limit, a Lorentzian excitation leads to a zero frequency noise spectral density $S=Ne^2\nu$, i.e. the same as the shot noise of a purely dc current $I=Ne\nu$. Thus, this time dependent periodic drive creates an out-of-equilibrium electron distribution function which leads to a charge transfer of $\overline{N}_e=N$ electrons per cycle in average with a variance $\Delta N_e^2 \sim N$ without any electron-hole excitation. The former results holds for a single channel conductor. For a tunnel junction of conductance $G=\gamma \frac{e^2}{h}$, the charge transfer is characterized by an average number of electron $\bar{N}_e= \gamma N$. A $50$~$\Omega$ matched tunnel junction corresponds to $\gamma \sim 500$.

The simplest way to generate pulses consists in adding a pure sine wave to a dc voltage such as $V(1+\cos (2\pi \nu t))$ and we will show in section \ref{sec:harm} that it is enough to see the electron-hole pairs reduction at quantized values of $eV/h\nu$ (see Fig.~\ref{fig:excess_noise}). However, a much richer waveform, which we present in section \ref{sec:biharm}, is the bi-harmonic drive:
\begin{equation}
V_{ac}(t)=V_{ac1} \cos(2\pi \nu t) + V_{ac2} \cos(4\pi \nu t+ \varphi).
\label{eqbiharm}
\end{equation}
\noindent characterized by three parameters $V_{ac1}, V_{ac2}$ and $\varphi$. By modifying them, we can control the out-of-equilibrium electron distribution function and thus the noise. For particular values of these parameters ($V_{ac1}=2V_{ac2}= h \nu$ and $\varphi=0$ or $\pi$) we can even approximate the ideal case of a sequence of Lorantzian pulses (see Fig.~\ref{fig_DeltaSii_th}(a)).

\begin{figure}[t]%
\includegraphics*[width=0.9\linewidth]{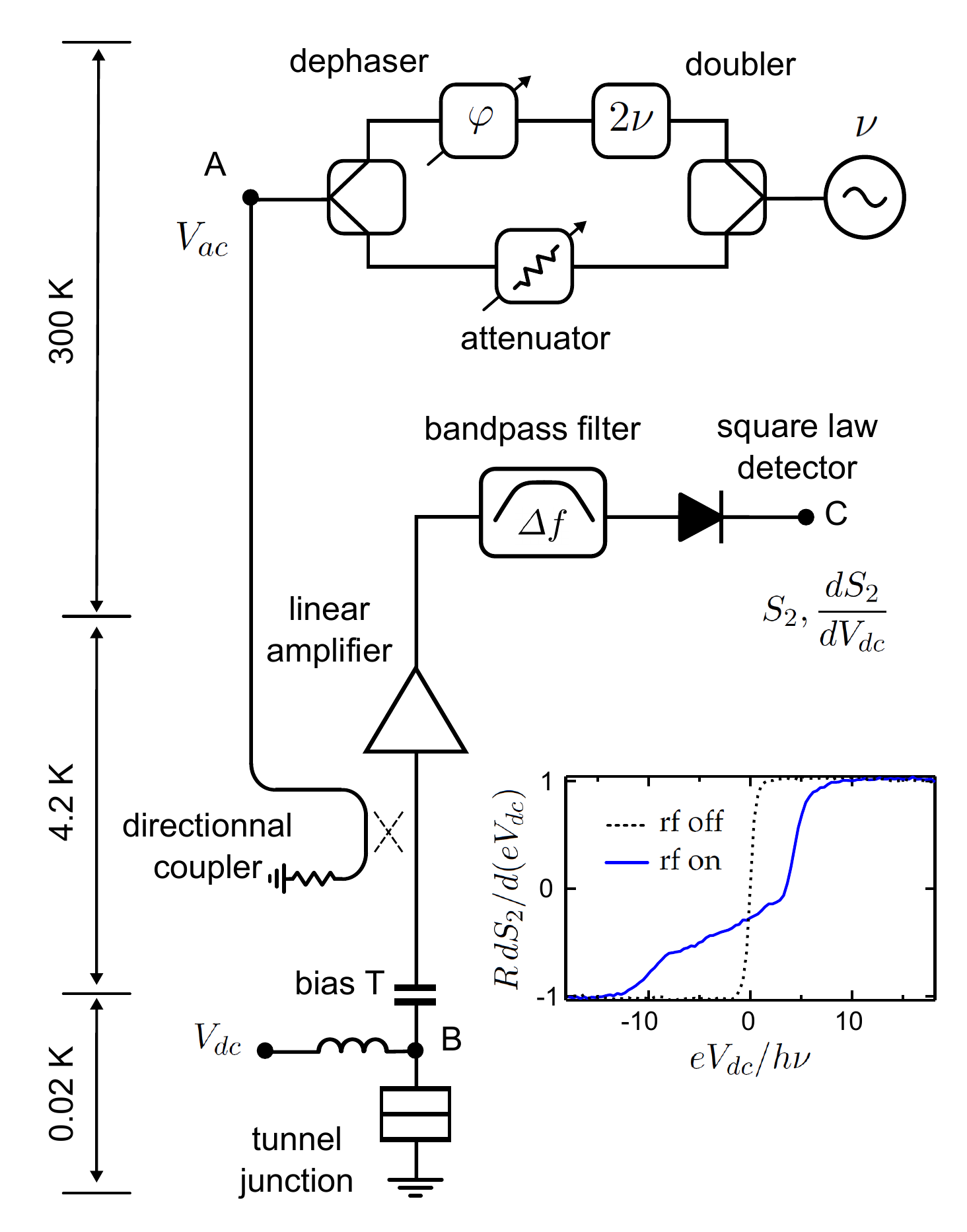}
\caption{Experimental setup for the dynamical charge transfert through a tunnel junction using harmonic and bi-harmonic pulses. Inset: normalized differential noise spectral density  with (blue line) and without (black dot line) microwave excitation \textit{vs.} normalized dc bias for $eV_{ac1}=2eV_{ac2}=5.4h\nu$ and $\varphi=0$.}
\label{fig_setup2}
\end{figure}

The setup to drive the junction and to detect the noise is sketched in Fig.~\ref{fig_setup2} and explained in \cite{Gabelli2013}. It allows the generation of two phase-locked sine waves with control of their two amplitudes and relative phase. The ac voltages experienced by the sample are calibrated with the photo-assisted noise with a single frequency excitation, as in \cite{Gabelli2008}.

\subsection{Harmonic pulses.}
\label{sec:harm}

\vspace{0.2cm}
\noindent Here, we report measurements of low frequency noise (1-80~MHz, $\ll k_BT/h=440$~MHz) generated by a tunnel junction of resistance $R=70$~$\Omega$ placed at very low temperature $T=27$~mK while being excited by a harmonic ac voltage at frequency $\nu=10$~GHz or $\nu=20$~GHz ($\gg k_BT/h$).

\begin{figure}[t]
\centering
\includegraphics[width=0.9\linewidth]{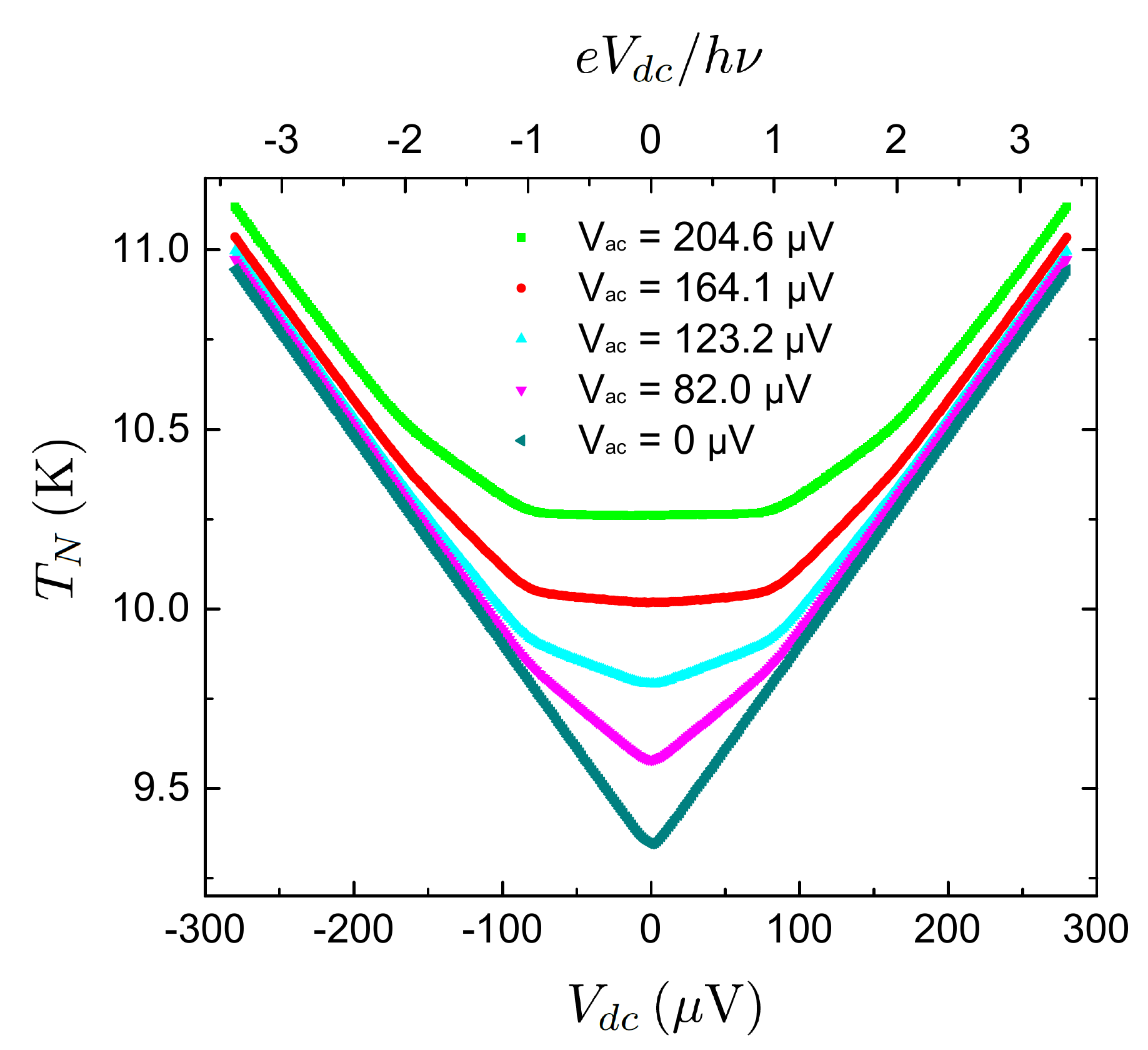}
\caption{Measured noise temperature as a function of dc voltage (lower axis) or reduced dc voltage $eV_{dc}/h\nu$ (upper axis) for various amplitudes of the ac excitation at frequency $\nu=20$~GHz. The theoretical fit, Eq.~(\ref{eq_PAN}), is indistinguishable from the experimental data. Taken from ref.~\cite{Gasse2013}.}
\label{fig:bruit_RF}
\end{figure}

Noise in the presence of an ac excitation at $\nu=20$~GHz is plotted on Fig.~\ref{fig:bruit_RF}. The dark green curve corresponds to $V_{ac}=0$. Fitting these data with Eq.~(\ref{eq:S(V,f=0)}) allows us to deduce the electronic temperature $T=27$~mK. In the presence of a finite $V_{ac}$ the zero frequency shot noise is given by Eq.~(\ref{eq5}) with:
\begin{equation}
\label{eq_PAN}
c_n=J_n\left(\frac{eV_{ac}}{h\nu} \right),
\end{equation}
where $J_n$ are the Bessel functions of the first kind. Our experimental data are indistinguishable from the theoretical predictions.

\begin{figure}[t]
\centering
\includegraphics[width=0.9\columnwidth]{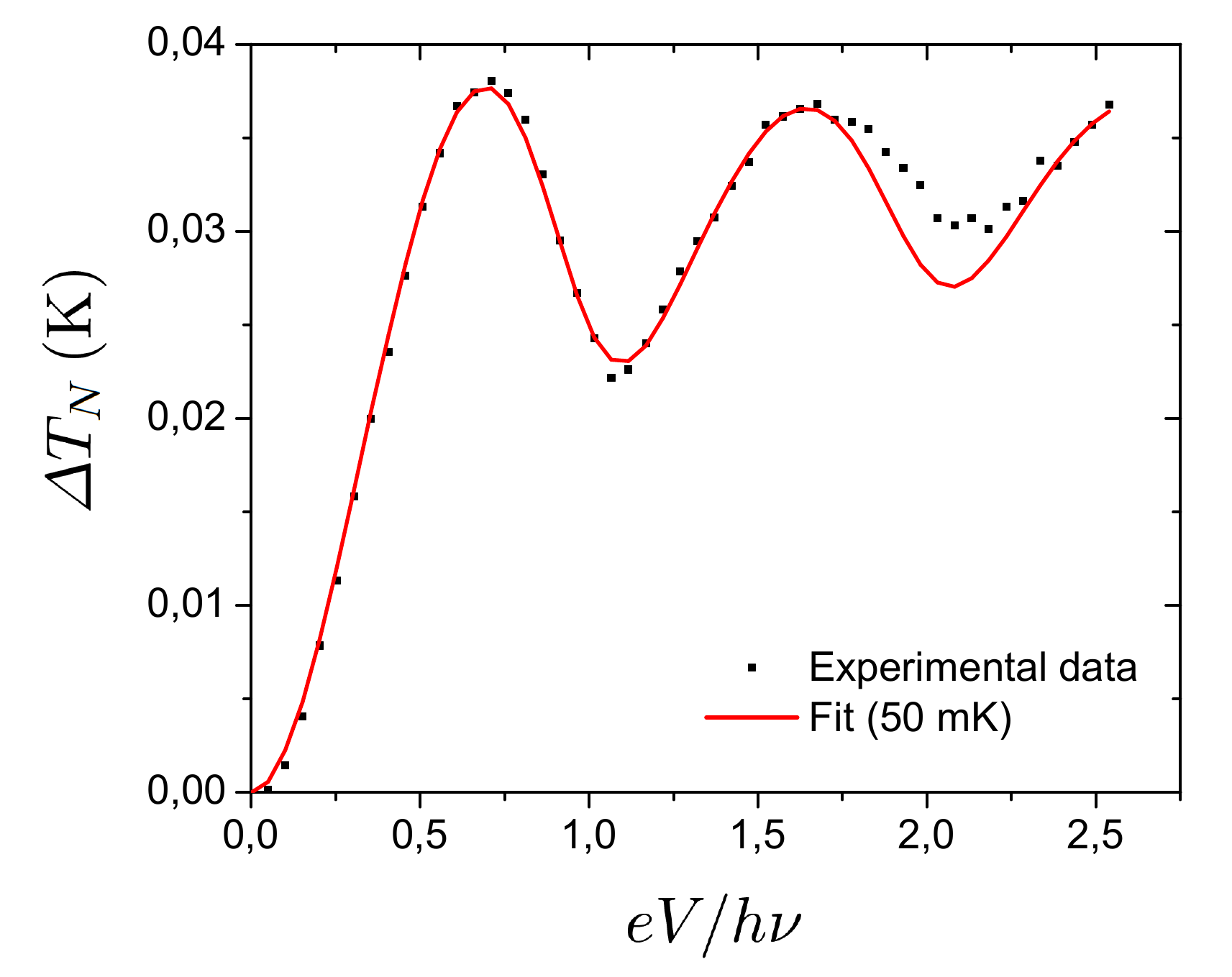}
\caption{Reduced excess noise $\Delta T_N = R\Delta S_V(V)/(2k_B)$ as a function of the amplitude $V$ of the time-dependent voltage excitation $V(1+\cos2\pi\nu t)$ with $\nu=20$~GHz applied on the sample. Black dots are experimental data, red line is theory for $T=50$mK.}
\label{fig:excess_noise}
\end{figure}

To reveal the quantum oscillations of the number of electron-hole excitations $N_{e-h}$, we excite the sample with a time-dependent voltage given by $V(t)=V(1+\cos2\pi\nu t)$, i.e. we sweep both the dc and ac voltages keeping  $V_{ac}=V_{dc}=V$. We measure the excess shot noise as a function of $V$, i.e.  $\Delta S_V(V)= S(V_{dc}=V_{ac}=V) - S(V_{dc}=V,V_{ac}=0)$, related to $N_{e-h}$ by Eq.~(\ref{eq:Neh}). The results for $\nu=20$~GHz as a function of $eV/h\nu$ are shown as black dots in Fig.~\ref{fig:excess_noise}. Oscillations of $\Delta S_V(V)$ with a period $h\nu/e$ are clearly visible on the experimental data. We observe that the oscillations are not exactly periodic. This is expected at finite temperature: when $T$ increases, the minima of $\Delta S_V$ move to higher voltage.

The theoretical expectations for $\Delta S_V(V)$ with a temperature $T=50$~mK is plotted as a red line in Fig.~\ref{fig:excess_noise}. Experimental data and theory match well at low voltage but the correspondence gets worse at higher voltage. This discrepancy is very well accounted for by the fact that the electron temperature depends on the ac bias and the slight mismatch between $V_{dc}$ and $V_{ac}$, see details in ref.~\cite{Gasse2013}.

\subsubsection{Discussion.}

\vspace{0.2cm}
\noindent The quantum oscillations in the shot noise of a tunnel junction shown on Fig.~\ref{fig:excess_noise} are directly proportional to the number of electron-hole pairs by $\Delta T_N = N_{e-h} (h\nu /2k_B)$. As expected for an harmonic ac excitation (see Fig.~\ref{fig:bruit_RF}), $N_{e-h}$ oscillates with a period $eV/h\nu$ but never reaches zero. One could perform the same experiment on any other coherent sample such as a metallic wire or quantum point contact and measure similar oscillations. However, the observed oscillations would have a lower amplitude because of the reduced Fano factor of these devices.

\subsection{Bi-harmonic pulses.}
\label{sec:biharm}


\vspace{0.2cm}
\noindent We are interested here in approximating the Lorentzian pulses with a bi-harmonic signal. In this experiment, the resistance of the $48$~$\Omega$ tunnel junction is close enough to the $50$~$\Omega$ input impedance of the detection setup to avoid reflection of the ac excitation ($|\Gamma|^2 < 10^{-3}$). The temperature noise of the amplifier is $T_A \simeq 7 \, \mathrm{K}$ and the measurement bandwidth $0.5 - 1.8 \, \mathrm{GHz}$. Before discussing the "purity" of the electron states generated by bi-harmonic pulses, we start by demonstrating how to dynamically control the distribution function of electrons in the contacts of a conductor.

\begin{figure}[t]%
\includegraphics*[width=0.9\linewidth]{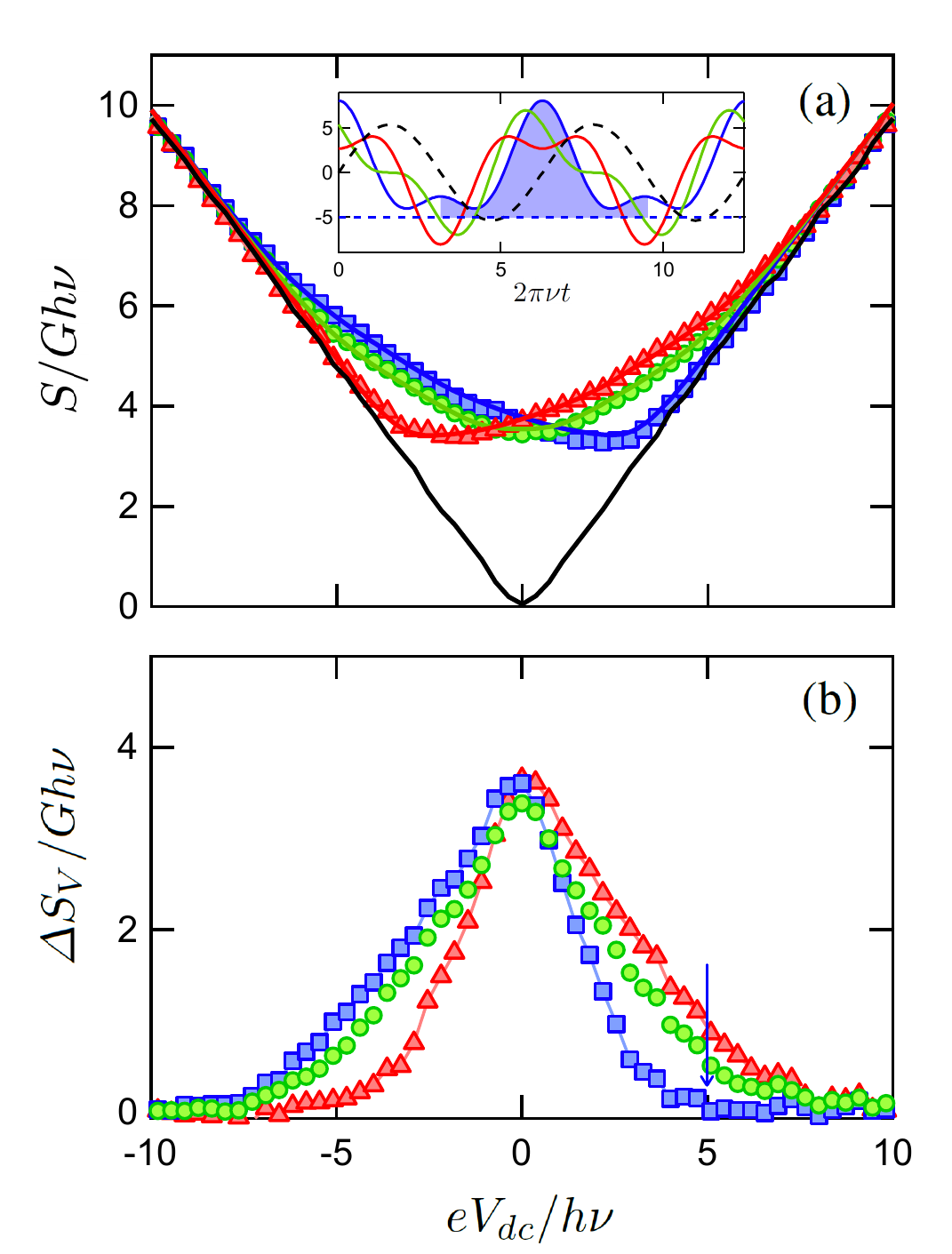}
\caption{ (a) Normalized biharmonic photon-assisted noise $S/Gh\nu$ as a function of normalized dc bias for $eV_{ac1}=5.4h\nu$. Blue square, green circle, red triangle symbols stand for data where $eV_{ac2}=2.7h\nu$ and phase shifts $\varphi= 0, \pi/2, \pi$. Black line: data for $V_{ac1}=V_{ac2}=0$ i.e. shot noise without any ac excitation. \textit{Inset}: $T$-Periodic sequence of bi-harmonic excitation for $\varphi=0$ (blue line), $\pi/2$ (green line) and $\pi$ (red line). The black dashed line corresponds to an harmonic signal. (b) Excess noise $\Delta S_{V}(V_{dc})=S(V_{dc},V_{ac1},V_{ac2})-S(V_{dc},V_{ac1}=0,V_{ac2}=0)$ normalized to $Gh\nu$.}
\label{fig_DeltaSii_exp}
\end{figure}

\subsubsection{Out-of-equilibrium electron distribution function.}

As shown in ref.~\cite{Gabelli2013}, the derivative of the noise $\partial S/ \partial eV_{dc}$ can be identified to the electron distribution function $f$:

\begin{equation}
 f(\epsilon_F+\xi) \simeq  \frac{1}{2}\left(1- \frac{1}{G} \frac{\partial S}{\partial eV_{dc}} \right)_{eV_{dc}=\xi},
\end{equation}

\noindent when $\xi \gg  h\Delta f, k_BT_{el}$. As shown in the inset of Fig.~\ref{fig_setup2}, this quantity exhibits a step-like function (black dotted line) at the Fermi level when the junction is dc voltage biased. However, this is not true anymore when the ac excitation is switched on (blue solid line in inset of Fig.~\ref{fig_setup2}). This shows how the distribution function of electrons can be controlled by the shape of the exciting waveform applied to the tunnel junction. The close relationship between shot noise and energy distribution function is also at the origin of Eq.~(\ref{eq5}), namely the expression of the out-of-equilibrium distribution function under ac excitation:

\begin{equation}
 f(\epsilon) =\sum_{n=-\infty}^{+\infty} \left|c_n \right|^2 \, f_{FD}(\epsilon+nh\nu),
\end{equation}

\noindent where $f_{FD}$ is the equilibrium Fermi-Dirac distribution. In the case of an harmonic excitation, one  observes in Fig.~\ref{fig:bruit_RF} discontinuities of $\partial S/ \partial eV_{dc}$ at bias $eV_{dc}= nh\nu$ with $n$ integer. For a bi-harmonic excitation, one observes a much richer structure. The interferences between multi-photon assisted processes at frequencies $\nu$ and $2 \nu$ induce interference patterns, as shown in Fig.~3 of ref.~\cite{Gabelli2013}. These interferences are theoretically described by the $c_n$ coefficients depending on the parameters $V_{ac1}$, $V_{a21}$ and $\varphi$ according to:

\begin{equation}
\label{eq_PANbiharm}
c_n=\sum_{m=-\infty}^{+\infty} J_{n-2m}\left( \frac{eV_{ac1}}{h\nu}\right)J_{m} \left( \frac{eV_{ac2}}{2h\nu}\right) \, e^{-im \varphi}.
\end{equation}

As a consequence, the minimum of the noise is not reached at $V_{dc}=0$ anymore. It is a first evidence that an asymmetry in the waveform enables to reduce the number of electron-hole pairs surrounding the transferred charge at $V_{dc} \neq 0$.

\subsubsection{Discussion.}

We have chosen a bi-harmonic drive with $V_{dc}\simeq V_{ac1}=2V_{ac2}$ in order to mimic a sequence of Lorentzian  pulses of width $\tau=\ln 2/(2\pi \nu)$ (see Fig.~\ref{fig_DeltaSii_th}(a)). Fig.~\ref{fig_DeltaSii_exp}(b) shows the normalized voltage excess noise $\Delta S_{V}/Gh\nu$ for $V(t)=V_{dc}+5.4 h\nu [\cos(2\pi \nu t)+0.5\cos(4\pi \nu t)]$. This quantity directly gives the mean number of electron-hole pairs $\bar{N}_{e-h}$ surrounding the electron excitation per cycle. In the case of $\varphi=0$ (blue square), we see in Fig.~\ref{fig_DeltaSii_exp}(b) a significant reduction of excess noise at $eV_{dc}/h\nu=5$ ($\bar{N}_{e-h}/\bar{N}_{e} \sim 1\%$, blue arrow) while the excitation depicted in the inset of Fig.~\ref{fig_DeltaSii_exp}(a) clearly shows a pulsed excitation signal (blue shadow) enabling a controlled charge transfer. It corresponds in the experiment to the transfer of $\bar{N}_e=\gamma N \sim 2500$ electrons with a cloud of $\bar{N}_{e-h} < 25$ electron-hole pairs. Table~\ref{tab} gives the average number of electron-hole pairs $N_{e-h}$ generated by harmonic and  bi-harmonic signals as a function of the number of transferred charges $N_{e}$. In the case of a bi-harmonic signal with $V_{dc}=V_{ac1}=2V_{ac2}$ and $\varphi=0,\pi$,  the relative size of the electron-hole cloud is always $N_{e-h}/N_{e}<0.7 \%$.
\begin{table}[b]
\begin{center}
\vspace{0.5cm}
  \begin{tabular}[htbp]{@{}lll@{}}
    \hline
    $N_{e}$ & $N_{e-h}^{harm}$ &  $N_{e-h}^{bi-harm}$ \\
    \hline
    1  & 0.028  & 0.007  \\
    2  & 0.039  & 0.012  \\
    3  & 0.044  & 0.019  \\
    4  & 0.047  & 0.022  \\
    5  & 0.05   & 0.019  \\
    \hline
  \end{tabular}
\end{center}
\caption{Number of the electron-hole pairs $N_{e-h}$ surrounding the transferred charge $N_e$ for an harmonic and a bi-harmonic drive depicted in Fig.~\ref{fig_DeltaSii_th}(a).}
  \label{onecolumntable}
  \label{tab}
\end{table}

The link between the shape of the ac bias $V_{ac}(t)$ and the number of created electron-hole pairs is well established at zero temperature, while experiments are of course performed at finite temperature. It is thus noteworthy to consider what is the number of electron-hole pairs that are thermally excited in an experiment. A naive answer could be the following: in the absence of ac voltage, the noise added by the finite temperature corresponds to the addition of electron-hole pairs. Thus the relevant excess noise is the thermal one $\Delta S_T$ defined in Eq.~(\ref{eq:deltaST}) that measures how much noise is added by the finite temperature. Supposing that a Lorentzian pulse does not add electron-hole pairs even at finite temperature, $N_{e-h}$ should be given by $N_{e-h}=\Delta S_T(f=0)/(Gh\nu)=\coth(Nh\nu/2k_BT)-1$. For $N=1$, $\nu=10$~GHz and $T=30$~mK one obtains $N_{e-h}\sim 10^{-6}$. According to this estimate, based on the zero frequency thermal excess noise, a voltage pulse on a coherent conductor may indeed be a quasi-pure single electron source. Things might however be not so simple. Let us now consider the harmonic excitation at finite temperature. The \emph{total} number of electron-hole pairs should be given by the difference between the noise at finite temperature in the presence of the ac excitation and the noise at zero temperature with no ac excitation (but the same dc voltage). According to our observations at $\nu=10$~GHz (not shown but similar to that of Fig.~\ref{fig:excess_noise} which corresponds to $\nu=20$~GHz), which matches the theoretical result very well, one obtains $N_{e-h}\simeq 0.056$ for $N=1$ , to be compared with the zero-temperature result $0.028$. The finite temperature thus doubles the number of electron-hole pairs ! The difference between the two approaches is that in the first case we consider the effect of the temperature at finite voltage on the noise without photo-excitation, which is insignificant, whereas in the second case we consider its effect on the photo-assisted noise at $eV=h\nu$, which is precisely where the thermal rounding of the noise is significant. In this case, the number of electron-hole pairs is of the order of $k_BT/h\nu$. In that condition, realizing a quasi-pure single electron source with voltage pulses is experimentally very demanding, and the utility of Lorentzian pulses rather than simple sine waves might be not so obvious. In any case, theoretical work is needed to understand the meaning of the number of electron-hole pair created by voltage pulses at finite temperature.

\section{Conclusion.}

\vspace{0.2cm}
\noindent We have shown that a simple voltage biased conductor can be used as a source for electron quantum optics in three regimes. First, we have reiterated that a dc voltage biased contact acts as a single electron turnstile where the electrons are regularly emitted from the macroscopic contact with a period $h/eV$. Second, we have detailed that, in the presence of harmonic pulses, the noise shows quantum oscillations that correspond to the passage of an integer number of electrons per period. This constitutes a first step towards an on-demand source of electrons, with a purity $1-N_{e-h}/N_{e} \sim 3 \%$. Third, with the use of a bi-harmonic excitation we have realized an electron source with a purity  $\sim 99 \%$. How finite temperature affects these numbers should however be theoretically explored.

\begin{acknowledgement}
We thank Marco Aprili, Wolfgang Belzig, Leonid Levitov and Mihajlo Vanevic for fruitful discussions. This work was supported by ANR-11-JS04-006-01, the Canada Excellence Research Chairs program, the MDEIE, the FCI, the NSERC, the FRQNT, the INTRIQ and the Canada Foundation for Innovation.
\end{acknowledgement}

\end{document}